\documentclass[aps,prl,twocolumn,showpacs,superscriptaddress, preprintnumbers,amsmath,amssymb]{revtex4-1}
\usepackage{graphicx}% Include figure files
\usepackage{epsfig}
\usepackage{dcolumn}% Align table columns on decimal point
\usepackage{bm}% bold math
\usepackage{longtable}
\usepackage{color}

\begin{document}

%\preprint{}

\title{Vortex lock-in transition and evidence for transitions among commensurate kinked vortex configurations in single-layered Fe arsenides}



\author{G. Li}
\affiliation{National High Magnetic Field Laboratory, Florida
State University, Tallahassee-FL 32310, USA}
\author{G. Grissonnanche}
\affiliation{National High Magnetic Field Laboratory, Florida
State University, Tallahassee-FL 32310, USA}
\author{B. S. Conner}
\affiliation{National High Magnetic Field Laboratory, Florida
State University, Tallahassee-FL 32310, USA}
\author{F.\ Wolff-Fabris}
\affiliation{Dresden High Magnetic Field Laboratory, Helmholtz-Zentrum Dresden-Rossendorf, D-01314 Dresden, Germany}
\author{C. Putzke}
\affiliation{Dresden High Magnetic Field Laboratory, Helmholtz-Zentrum Dresden-Rossendorf, D-01314 Dresden, Germany}
\author{N.\ D.\ Zhigadlo}
\affiliation{Laboratory for Solid State Physics, ETH
Z\"{u}rich, CH-8093 Z\"{u}rich, Switzerland}
\author{S.\ Katrych}
\affiliation{Laboratory for Solid State Physics, ETH
Z\"{u}rich, CH-8093 Z\"{u}rich, Switzerland}
\affiliation{Institute of Condensed Matter Physics, EPFL CH-1015 Lausanne, Switzerland}
\author{Z.\ Bukowski}
\affiliation{Laboratory for Solid State Physics, ETH
Z\"{u}rich, CH-8093 Z\"{u}rich, Switzerland}
\author{J.\ Karpinski}
\affiliation{Laboratory for Solid State Physics, ETH
Z\"{u}rich, CH-8093 Z\"{u}rich, Switzerland}
\affiliation{Institute of Condensed Matter Physics, EPFL CH-1015 Lausanne, Switzerland}
\author{L. Balicas} \email{balicas@magnet.fsu.edu}
\affiliation{National High Magnetic Field Laboratory, Florida
State University, Tallahassee-FL 32310, USA}

%\thanks{}

\date{\today}

\begin{abstract}
We report an angle-dependent study of the magnetic torque $\tau(\theta)$ within the vortex state of
single-crystalline LaO$_{0.9}$F$_{0.1}$FeAs and SmO$_{0.9}$F$_{0.1}$FeAs as a function of both
temperature $T$ and magnetic field $H$. Sharp peaks are observed at a critical angle $\theta_c$ at either side of $\theta=90^{\circ}$, where
$\theta$ is the angle between $H$ and the inter-planar \emph{c}-axis. $\theta_c$ is interpreted as the critical depinning angle where
the vortex lattice, pinned and locked by the intrinsic planar structure, unlocks and acquires a component
perpendicular to the planes. We observe a series of smaller replica peaks as a function of $\theta$ and as $\theta$ is
swept away from the planar direction. These suggest commensurability effects between the period of the vortex lattice and the inter-planar distance leading to additional kinked vortex configurations.
\end{abstract}

\pacs{74.25.-q, 74.25.Uv, 74.25.Wx, 74.70.Dd}
%\keywords{}

\maketitle

%\section{Introduction}
The recently discovered superconducting (SC) Fe oxypnictides
\cite{discovery,chen} are characterized by high upper critical fields \cite{Hc2},
or small superconducting coherence lengths, relatively small anisotropies \cite{anisotropy} and apparently, high superconducting critical currents \cite{Jc}. Nevertheless, to date
the important properties of their mixed state have been little explored.
In the \emph{R}LaFeAsO$_{1-x}$F$_x$ (where \emph{R} is a rare-earth element) or 1111 compounds,
it was found that the main contribution to the pinning of the vortex lines comes from the collective pinning by the dopant atoms, whose
local density variations would seem to lead to strong pinning \cite{beek}.

These compounds are composed of alternating layers of SC and non-SC material, which leads to anisotropic electronic properties both above and below the SC transition temperature $T_c$. For strongly coupled layers, the superconducting pair amplitude is just weakly modulated by the discrete structure. Thus, the coherence length in the direction perpendicular to the planes $\xi_{c}$ is much larger than the characteristic inter-planar distance $c$, and the system is well described by a continuum model with an anisotropic mass tensor \cite{GL}. If the coupling between the SC layers is weak, this pair amplitude is large only on the SC planes, and one could have a situation where $\xi_{c} < c$, and where the discrete layers would be coupled by the Josephson tunneling \cite{LD}. In the strong coupling case, a magnetic field parallel to the SC layers generate vortices having a core region of suppressed SC of size $\sim \xi_{c} \times \xi_{ab} $ spanning several layers. While in the weak coupling limit the core of the vortex line fits between the planes, where the pair amplitude is small. If one increases the component of the field $H_{\bot}$ perpendicular to the layers, the vortex lattice remains in a ``locked-in" state with the magnetic flux lines trapped by the layers until $H_{\bot}$ creates normal cores on them. Above this threshold field, the tilted flux lines would pierce the layers forming a staircase pattern of kinked vortices \cite{LD}.

Remarkably, for fields aligned along the \emph{ab}-plane and upon cooling below a certain temperature $T^{\star}$, a recent electrical-transport study \cite{mollNM} in both the SmFeAsO$_{1-x}$F$_x$ and the LaFeAsO$_{1-x}$F$_x$ compounds found a dramatic \emph{decrease} in critical current densities ($J_c$) for currents flowing along the \emph{c}-axis. $T^{\star}$ corresponds to the temperature at which the inter-planar coherence length matches half of the \emph{c}-axis, implying that the vortices fit in between the SC layers. This observation is interpreted as evidence for a transition from well-pinned and slow moving Abrikosov vortices at higher temperatures, to weakly pinned and therefore fast flowing Josephson-vortices at $T \leq T^{\star}$. This contrasts with the small SC anisotropy of these compounds. The phase core of such Josephson vortices avoids the large SC order-parameter in the FeAs layers, which effectively generates a potential barrier impeding the flow of Josephson vortices perpendicularly to the layers but not along them as seen in Ref. \cite{mollNM}, creating a situation analogous to the ``locked-in" state. When the field is slightly misaligned with respect to the \emph{ab}-plane, Refs. \onlinecite{Jc, mollNM} find a dramatic enhancement in $J_c$ due to strong pinning of the metallic cores of Abrikosov vortices. These observations imply the existence of a lock-in transition from a lattice of Josephson-vortices for fields along a planar-direction to an Abrikosov vortex-solid as the field rotates towards the \emph{c}-axis.
\begin{figure}[htb]
\begin{center}
\epsfig{file= 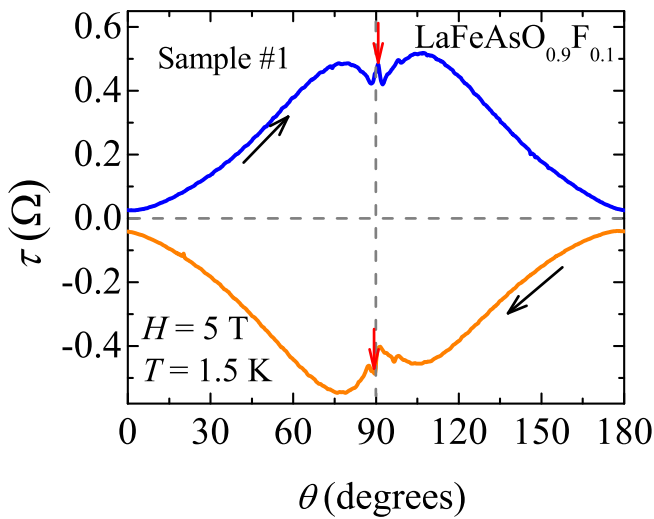, width = 8.6 cm} \caption {(Color online) Magnetic torque $\tau$ for a
LaO$_{0.9}$F$_{0.1}$FeAs single crystal for increasing (blue line) and
decreasing (orange line) angle ($\theta$) sweeps, at $H=5$ T and $T=1.5$ K. $\theta=0^{\circ}$ corresponds to fields ($H$) along the inter-planar direction. Sharp spikes indicated by red arrows, are observed at either side of $\theta = 90^{\circ}$.}
\end{center}
\end{figure}
\begin{figure}[htb]
\begin{center}
\epsfig{file= 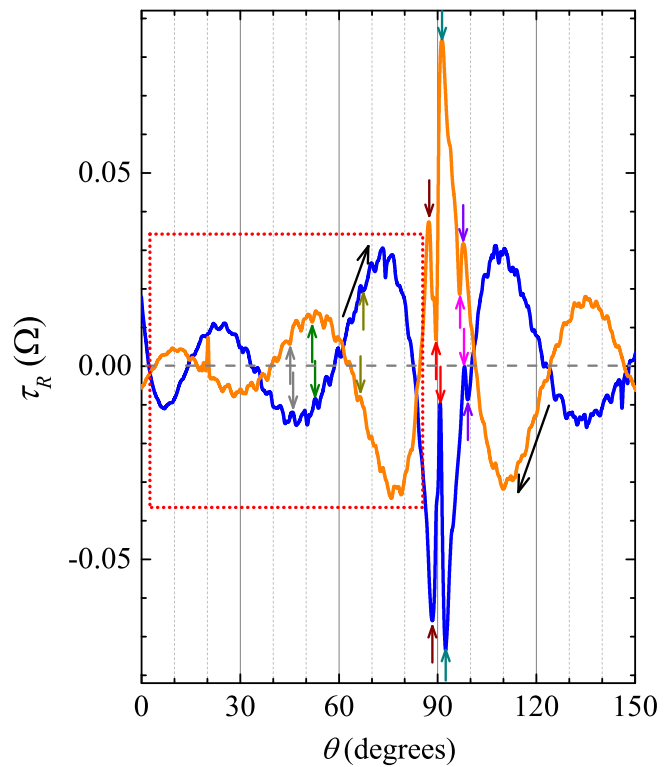, width = 8.6 cm} \caption {(Color online) From the traces in Fig. 1, the subtraction of a ninth-degree polynomial leads to a residual term, $\tau_{R}(\theta)$ containing sharp pronounced features as well as smaller nearly oscillatory structures. Each pair of colored arrows indicate a pair of corresponding features such as sharp peaks or dips which are seen on each trace, but which are slightly displaced in angle. The separation between the red arrows yields $2\theta_c \simeq 2.36^{\circ}$.  Square indicates the angular region chosen to analyze the fine structure displayed below in Fig. 3 (a).}
\end{center}
\end{figure}

In order to provide thermodynamic evidence for such a transition, here we present a detailed angle-dependent study of the magnetic torque $\overrightarrow{\tau}(\theta) = \mu_0 \overrightarrow{m} \times \overrightarrow{H} = \mu_0 mH \sin(\theta) $  ($\overrightarrow{m}$ is the magnetization of the sample and $\theta$ is the angle between $H$ and the c-axis) in single crystals of LaFeAsO$_{0.9}$F$_{0.1}$ and SmFeAsO$_{0.9}$F$_{0.1}$. For both compounds, we observe pronounced peaks in $\tau(\theta)$ for $\theta$ very close to $90^{\circ}$ and as the field is rotated away from the \emph{ab}-plane. The torque increases fast as $\theta$ is moved away from $90^{\circ}$, until a critical value for the component of the field perpendicular to the layers $H^c_{\bot}$ is reached, indicating that the vortex lines were originally locked in-between the superconducting planes. At $H^c_{\bot}$ the vortex lattice would have to undergo a transition towards a kinked vortex structure. We observe a succession of smaller replica peaks in $\tau(\theta)$ as $\theta$ is further displaced away from the \emph{ab}-plane suggesting a cascade of transitions among kinked vortex configurations. Details concerning samples and experimental set-up can be found in Supplemental Material \cite{supplemental}.
\begin{figure}[htb]
\begin{center}
\epsfig{file= 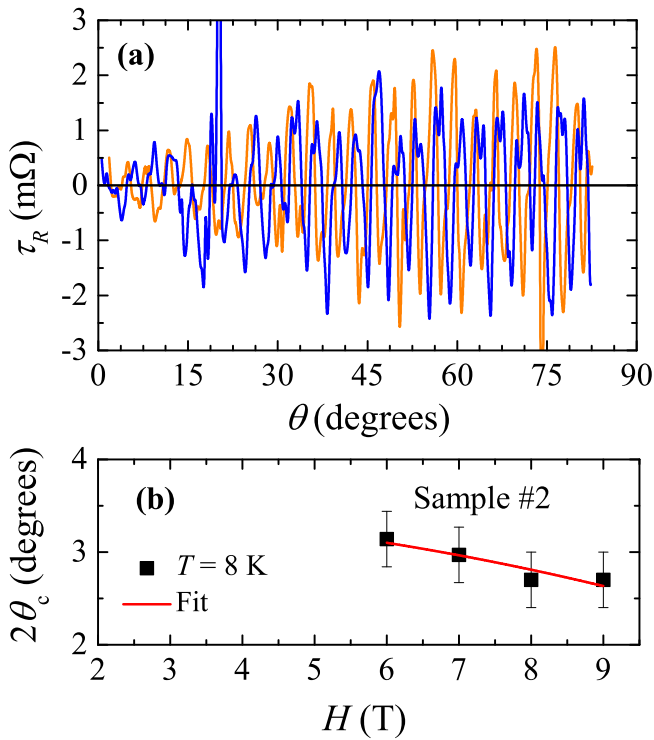, width = 8.6 cm} \caption {(Color online) (a) Oscillatory component $\tau_{R}(\theta)$ superimposed onto the $\tau(\theta)$ traces shown in Fig. 1 (a).
$\tau_{R}(\theta)$ was obtained by fitting $\tau(\theta)$ to a polynomial within a limited angular range. As in Fig. 1 (a) blue and orange lines depict traces for field-increasing and decreasing sweeps, respectively. Notice i) the reproducibility of the oscillatory structure and ii) how their amplitude decreases as $H$ is rotated towards the \emph{c}-axis ($\theta = 0^{\circ}$). Each oscillation in a given trace is slightly displaced in angle relative to its counterpart on the other trace. (b) Field dependence of twice the critical lock-in angle 2 $\theta_c$ for a second LaFeAsO$_{0.9}$F$_{0.1}$ single-crystal at $T=8$ K.}
\end{center}
\end{figure}

Figure 1 (a) shows the magnetic torque as a function of the angle $\theta$ measured in a LaO$_{0.9}$F$_{0.1}$FeAs single crystal, and respectively for increasing and decreasing $\theta$ sweeps at $T= 1.5$ K and $H = 5 $ T. A large hysteresis is observed between increasing and decreasing angle traces which, within the Bean model, \cite{bean} is proportional to $J_c$. Here, we will not discuss this aspect. Instead, we focus on the observed sharp peaks for $\theta$ very close to the inter-planar direction, which for each trace is placed respectively at $\theta_c \sim (90 \pm 1)^{\circ}$. Very similar features were previously observed in the cuprates \cite{intrinsic_pinning_cuprates}, and in less anisotropic materials \cite{intrinsic_pinning_others} for $H$ along a planar direction, and in both cases attributed to intrinsic pinning.
At higher $T$s, in addition to the contribution from the various vortex pinning mechanisms leading to the
hysteresis observed here, $\tau(\theta)$ is dominated by the orbital contribution of the vortex lines analyzed by us by using a modified version of the Kogan equation, with the resulting SC parameters (such as the SC anisotropy $\gamma \sim 10$) and analysis given elsewhere \cite{gang}. The critical lock-in depinning angle in anisotropic and Josephson coupled layered superconductors is predicted to follow \cite{LD, bulaevskii}:
\begin{equation}
    \theta_c \simeq 2 \frac{L_z}{L_y} \frac{H_{c1}^c}{ H} \frac{\ln(\alpha d/\xi_{ab} \gamma)}{\ln(\lambda_J/ \xi_{ab})},
    \label{lock_in_angle}
    \end{equation}
    where $L_z/L_y \sim 10^{-1}-10^{-2}$ is the ratio between the thickness and the length of the single crystal (of dimensions $ \sim 5 \times 60 \times 75$ $\mu$m$^3$), $\alpha$ is constant of the order of the unity, $H_{c1}^c$ is the SC lower critical-field for fields along the c-axis, $d$ is the inter-lattice spacing, $\gamma= \left(m_c/m_{ab}\right) $ is the GL mass anisotropy, with $\xi_{ab}$ being the in-plane coherence-length, and $\lambda_J = \gamma d$ is the Josephson length, respectively. For mildly anisotropic cuprates such as YBa$_2$Cu$_3$O$_{7-\delta}$, in the range of fields used for our measurements, this expression predicts $\theta_c(T) \sim 0.1^{\circ} - 1^{\circ}$ with respect to the \emph{ab}-plane \cite{LD}, as seen by us in  LaO$_{0.9}$F$_{0.1}$FeAs.  Nevertheless, given their relatively small anisotropy \cite{anisotropy} it remains unclear if Eq. \ref{lock_in_angle} is appropriate for the Fe arsenides.
\begin{figure}[htb]
\begin{center}
\epsfig{file= 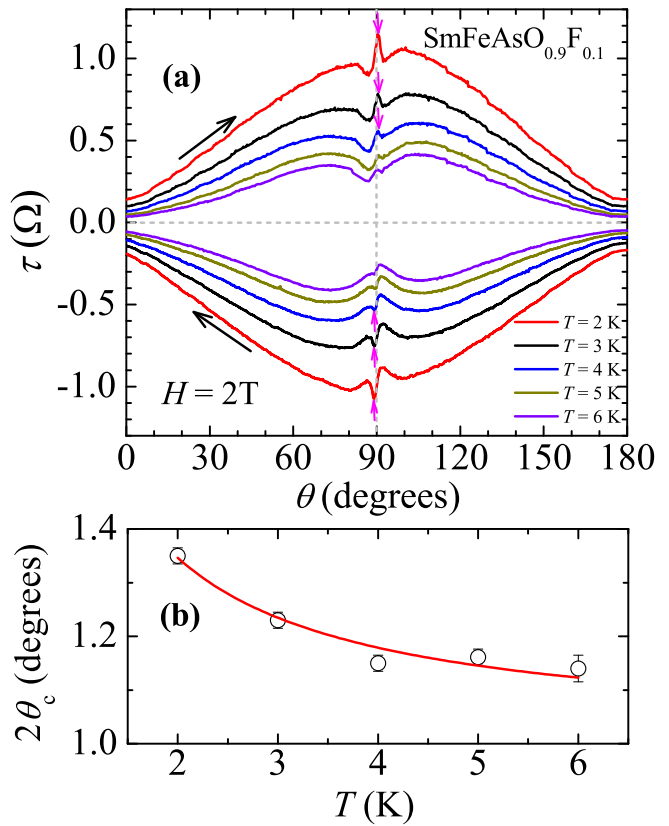, width = 8.6 cm} \caption {(Color online) (a) Magnetic torque $\tau(\theta)$ measured under a field $H = 2$ T for a
SmFeAsO$_{0.9}$F$_{0.1}$ single-crystal as a function of the angle $\theta$ and for several temperatures.
Included are traces for increasing and decreasing angle sweeps, respectively.
Sharp peaks are observed in $\tau(\theta)$ at an angle $\theta_c$ located at either side of $\theta = 90^{\circ}$.
(b) Difference in angle $2\theta_c$ between both peaks as a function of temperature. Red line is a guide to the eyes.}
\end{center}
\end{figure}

Fig. 2 shows the residual torque signal $\tau_R (\theta)$ after fitting both $\tau (\theta)$ traces in Fig. 1 to a polynomial and subtracting it from the original $\tau (\theta)$ data. As seen, very sharp and pronounced features are observed in both traces, and their amplitude decreases as the angle is swept towards $\theta = 0^{\circ}$, as if they were replicas of the most pronounced features seen at $\theta_c$. As clearly indicated by the colored arrows, features seen on one of these traces are also observed on the other one as if reflected on a mirror but slightly shifted in angle. These reproducible and sharp features seen in the hysteretic response can only correspond to pronounced changes in the pinning forces and/or concomitant pinning mechanisms.

In order to clearly expose the finer structure seen in $\tau (\theta)$ as $\theta \rightarrow 0^{\circ}$, Fig. 3 (a) displays $\tau_R (\theta)$ on a limited angular range, i.e. $0^{\circ} \leq \theta \leq 80^{\circ}$ after the subtraction of the slowly varying background in Fig. 2. Again, each oscillation seen in one of the traces is replicated by a similar oscillation on the other one although slightly displaced in angle.  $\theta_c$ on the other hand, should follow an $H^{-1}$ dependence. To check this, we extracted the $H$ dependence of $\theta_c$ as shown in the inset in Fig. 3 (b) for a second LaFeAsO$_{0.9}$F$_{0.1}$ single crystal (dimensions $\sim 8 \times 75 \times 95$ $\mu$m$^3$) at $T = 8$ K. $\theta_c$ does decrease slightly with $H$, i.e. red line is a fit to $\theta_c \propto H^{-1}$. When compared to sample \# 1 the comparatively larger values of $\theta_c$ in this sample could be attributed to i) geometrical factors ($\frac{L_z}{L_y}$ is $\sim 20$ \% larger) or ii) a sample-dependent interplay between intrinsic and point pinning. \cite{kugel} Notice that $\theta_c \sim 1^{\circ}$ with respect to the \emph{ab}-plane is considerably larger than $\theta_c \sim 0.1^{\circ} - 0.15^{\circ}$ required to observe the free flow of vortices in Ref. \onlinecite{mollNM}. Perhaps, at angles $\theta$ between $\sim 0.15^{\circ}$ and $\sim 1^{\circ}$ the interaction between Josephson vortices and planar defects would create kink and anti-kink pairs contributing to pinning, although a true a thermodynamic transition towards a kinked vortex lattice might require a $\theta_c \sim 1^{\circ}$.
\begin{figure}[htb]
\begin{center}
\epsfig{file= 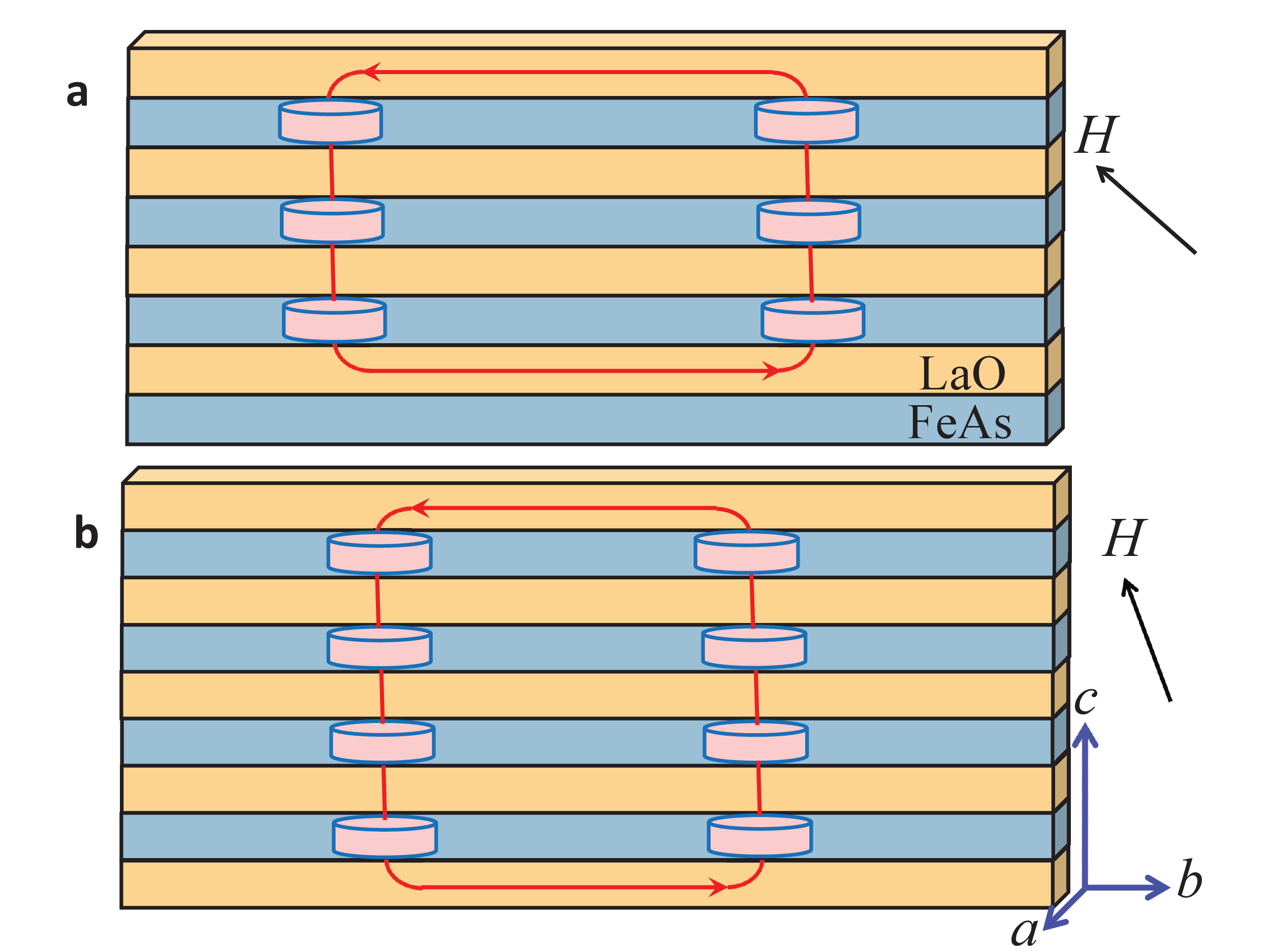, width = 8.6 cm} \caption {(Color online) a. Sketch depicting a single flux line in a layered superconductor for fields slightly tilted away with respect to the planar direction. The vortex line can be understood as an array of Abrikosov vortices (pink pancakes) piercing single superconducting layers which are interconnected by Josephson strings (red lines). At a small angle a tilted vortex line is composed of the two-dimensional core of the Abrikosov vortex and the phase core of the Josephson string. b. By rotating the field towards the c-axis, one shortens the Josephson strings, increases their inter-layer separation, and increases the density of Abrikosov vortices.}
\end{center}
\end{figure}

To demonstrate that the above observed features are intrinsic to the \emph{RE}FeAsO$_{1-x}$F$_x$ series, we have also measured $\tau(\theta)$ for a SmFeAsO$_{0.9}$F$_{0.1}$ single crystal under a field $H=2$ T and for several temperatures, as shown in Fig. 4 (a). Sharp peaks near to $\theta = 90^{\circ}$, as previously observed in the La compound, are also seen here. Their $T$ dependence is given in Fig. 4 (b), where $\theta_c$ is seen to follow a $\sim T^{-1}$ dependence. This is consistent with Eq. (1), since $\theta_c$ should depend on temperature as $H_{c1}^c \propto \lambda^{-2} \propto n_s$ where $n_s$ is the superfluid density, thus increasing  as $T$ is lowered. Notice that $\theta_c$ decreases quickly as $T$ increases and becomes basically unobservable above $T = 8 $ K. As discussed above, and in Ref. \onlinecite{mollNM}, the inter-layer coherence length $\xi_c$ increases with $T$ and above a temperature $T^{\star} \sim T_{c}/2$, $\xi_c$ becomes larger than $c/2$. For $H$ applied along a planar direction this leads to a transition from Josephson to pinned Abrikosov vortices, explaining the disappearance of the lock-in transition as $T$ increases. As shown in the Supplemental Material \cite{supplemental}, a fine periodic structure superimposed onto $\tau (\theta)$ is also observed for this compound and it disappears as either $H$ or $T$ increases, as expected from the above discussion.

In very anisotropic superconductors the SC order-parameter is expected to exhibit strong oscillations with the period $c$. Fields along a planar direction, forces the vortex lattice to accommodate itself to the layered structure so the vortex cores come to lie in between the superconducting layers. This scenario is expected to be valid in the limit $\xi_{c} \lessapprox c/2$, which in the present case has been shown to occur when $T \leq T^{\star}$. \cite{mollNM} But, as the field is tilted, vortices having a phase core of dimension $\xi_{c} \times \xi_{ab}$ are expected to undergo a kinked structure through a first-order phase transition \cite{bulaevskii} (lock-in transition) when crossing the superconducting planes and develop normal cores of dimension $\xi_{ab}^2$ within the planes. The sharp anomalies seen for fields nearly along a planar direction are thermodynamic evidence for a sharp change in pinning mechanisms thus being consistent with this scenario.

Furthermore, through electrical transport measurements we have found that as a field precisely aligned along the \emph{ab}-plane is varied, both the resistivity and $J_c$ display marked oscillations \cite{moll_to_be_published}. The maxima in critical current density is observed where the resistivity displays a minima, while its minima is observed at the maxima of the resistivity (maxima of dissipation), indicating that the vortex-lattice undergoes a series of transitions among commensurate and therefore locked-in configurations of Josephson vortices. The periodicity of the oscillations in the resistivity are consistent with so-called ``lock-in" oscillations observed in magnetization measurements \cite{deligiannis} in high purity YBa$_2$Cu$_3$O$_{7-\delta}$ single-crystals for fields parallel to the layers. These were attributed to commensurability effects between inter-layer spacing $d$ and the average vortex distance $\ell=nd$ along the inter-planar direction for a triangular vortex lattice. In the case discussed here, one also expects to induce a series of transitions among commensurate vortex configurations as the intensity of the field along the SC planes is varied by varying the angle. However, our situation is far more complex since for $\theta > \theta_c$ one stabilizes kinked vortex structures, which as illustrated by Fig. 5 would be composed of segments of Josephson vortices in between the superconducting planes, whose separation is adjusted discontinuously in order to match a few lattice spacings $\ell = nd$ \cite{kugel}, and become Abrikosov vortices when piercing the superconducting planes. By tilting the field away from the superconducting planes, one i) shortens the length of the Josephson segments, ii) increases their separation $\ell$ and iii) increases the density of Abrikosov vortices piercing the superconducting planes (See Figs. 5 a and 5 b). Our periodic structure could correspond to evidence for a cascade of transitions among kinked but commensurate vortex configurations \cite{doniach}.

We acknowledge useful discussions with P. J. W. Moll. L.~B. is supported by DOE-BES through award DE-SC0002613.
The NHMFL is supported by NSF through NSF-DMR-0084173 and the State of Florida.
Work at ETH was supported by the SNSF and the NCCR program MaNEP.
J.K. and S.K. acknowledge support from ERC Project Super Iron.

%\bibliography{balicasbib}
\end{document}